\documentclass[conference,compsoc]{IEEEtran}

\usepackage{epsfig}
\usepackage{enumitem}
\usepackage{multirow}
\usepackage[bookmarks=false]{hyperref}
\usepackage{graphicx}
\usepackage{booktabs}
\usepackage{microtype}
\hypersetup{
    colorlinks=true,
    linkcolor=blue,
    filecolor=magenta,      
    urlcolor=blue,
}

\title{A Comparative Study of Existing and New Deep Learning Methods \\
for Detecting Knee Injuries using the MRNet Dataset}

\author{\IEEEauthorblockN{David Azcona, Kevin McGuinness and Alan F. Smeaton}
\IEEEauthorblockA{Insight Centre for Data Analytics\\
Dublin City University\\
Glasnevin, Dublin 9, Ireland\\
Email: david.azcona2@mail.dcu.ie}}

\begin{document}

\maketitle


\begin{abstract}

This work presents a comparative study of existing and new  techniques to detect knee injuries by leveraging Stanford's MRNet Dataset. All approaches are based on deep learning and we explore the comparative performances of transfer learning and a deep residual network trained from scratch. We also exploit some characteristics of Magnetic Resonance Imaging (MRI) data by, for example, using a fixed number of slices or 2D images from each of the axial, coronal and sagittal planes as well as combining the three planes into one multi-plane network. Overall we achieved a performance of 93.4\% AUC on the validation data by using the more recent deep learning architectures and data augmentation strategies. More flexible architectures are also proposed that might help with the development and training of models that process MRIs. We found that transfer learning and a carefully tuned data augmentation strategy were the crucial factors in determining best performance.


\end{abstract}


\section{Introduction} \label{sec:intro}

Deep learning and convolutional neural networks trained using back propagation have revolutionised many aspects of computer vision, especially within the last decade, though it has not made  those computer vision techniques obsolete \cite{10.1007/978-3-030-17795-9_10}. In healthcare, areas such as traumatology are being positively impacted by these advances in vision especially when it comes to diagnostic imaging of x-rays, scans, etc. Analysing huge amounts of patient data such as scans or Magnetic Resonance Imaging (MRIs) are repetitive tasks for human experts and thus prone to errors for the human eye as a result of fatigue. Thus any assistance that automated processing can offer is to be welcomed. 

Deep neural networks and the ability to automatically analyse and extract non-linear patterns from vast amounts of image data are allowing radiologists to focus on cases that are unusual outliers or at the edges. This means that routine cases can be processed automatically and often with greater accuracy than human diagnosis because of the consistency in the analysis \cite{deo2015machine} and thus we can leverage this technology for everybody's benefit.

\begin{figure}[htp]
\centering
\includegraphics[width=.5\textwidth]{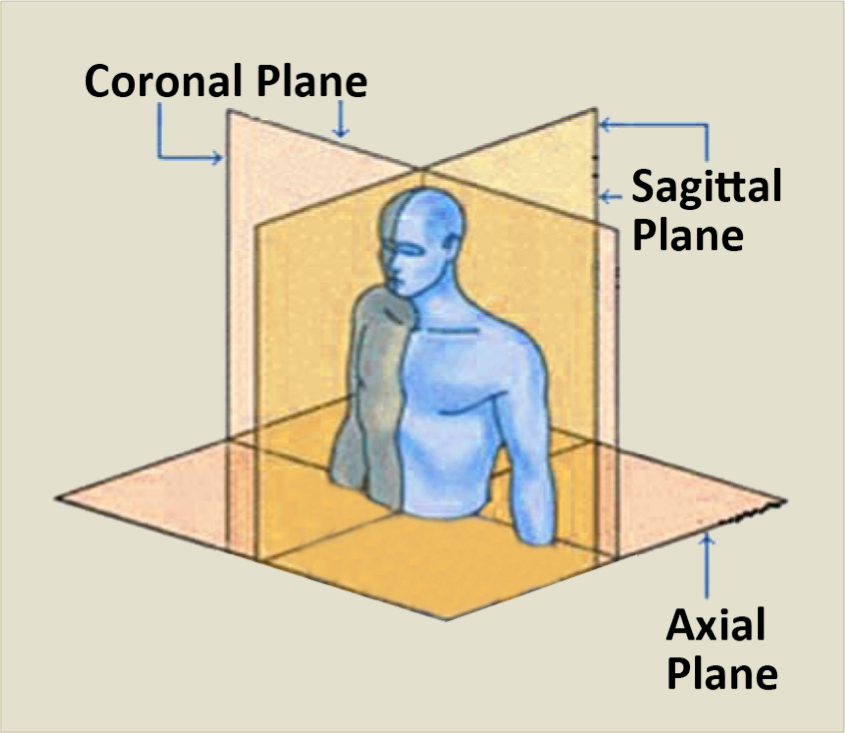}
\caption{MRI Image Reconstruction Planes~\cite{radiologyrounds}}
\label{fig:planes}
\end{figure}

Magnetic resonance imaging is a technique used in radiology for creating images of parts of the human body. These are created using large and powerful magnets and have widespread use for imaging almost all parts of the body but especially for imaging the brain, spinal cord, bones and joints, the heart and other internal organs. MRI scanning is a progressive operation that generates multiple two-dimensional cross-sections  or ``slices'' of tissue and from these we can generate three-dimensional reconstructions. MRI scans are usually created from three different orientations or planes, namely from the side (sagittal plane), front (coronal plane), and top-down (axial plane). Figure~\ref{fig:planes} shows a visual description of these planes. MRIs in three planes are thus more complex as a form of diagnostic imaging than, say, X-rays, which are just two dimensional.

In this paper we apply contemporary approaches from deep learning to a public dataset of knee MRI images, the MRNet dataset \cite{bien2018deep}, in order to automatically predict a diagnosis of whether the patient has a ligament tear or not. Some of these are existing and described in \cite{bien2018deep} while others are new and introduced here. The rest of this paper describes a selection of deep learning approaches we have implemented on this dataset. Our results are, at the time of writing, best-in-class and the paper discusses why our approaches work well on this particular kind of classification task. The lessons to be learned are:

\begin{enumerate}[leftmargin=*,label=\roman*)]   
\item transfer learning, even from a very different domain (natural images), plays a crucial role in regularising the network and preventing overfitting to the smaller MRNet dataset;
\item too much or too little data augmentation can be detrimental to performance, and a carefully tuned augmentation policy can give important performance benefits; and
\item modern network architectures with residual connections improve performance with higher AUC (Area Under the Curve) when combined with transfer learning and careful augmentation.
\end{enumerate}

The rest of the paper is organised as follows. In the next section we introduce the dataset and an implementation to explore the imagery. Then we discuss related work and the proposed MRNet architecture from Stanford researchers. This is followed by a section that gives details of each of the proposed architectures. We then outline experimental results using the AUC evaluation metric for each of our proposed architectures followed by the conclusions and future work.


\section{The MRNet Dataset} \label{sec:dataset}

The MRNet dataset \cite{bien2018deep} has been made publicly available by researchers from Stanford University in 2018. It consists of 1,370 knee MRI exams performed at the Stanford University Medical Center. The dataset contains MRIs for:
\begin{enumerate}[leftmargin=*,label=\roman*)]   
\item 1,104 (80.6\%) abnormal exams
\item 319 (23.3\%) ACL (anterior cruciate ligament) tears
\item 508 (37.1\%) meniscal tears
\end{enumerate}
Labels for each MRI were obtained through manual extraction from clinical reports on the patients. The dataset is very imbalanced in favour of labels with injuries as patients who undergo an MRI are more likely to suffer a knee injury. 

\begin{figure*}[htp]
\centering
\includegraphics[width=\textwidth]{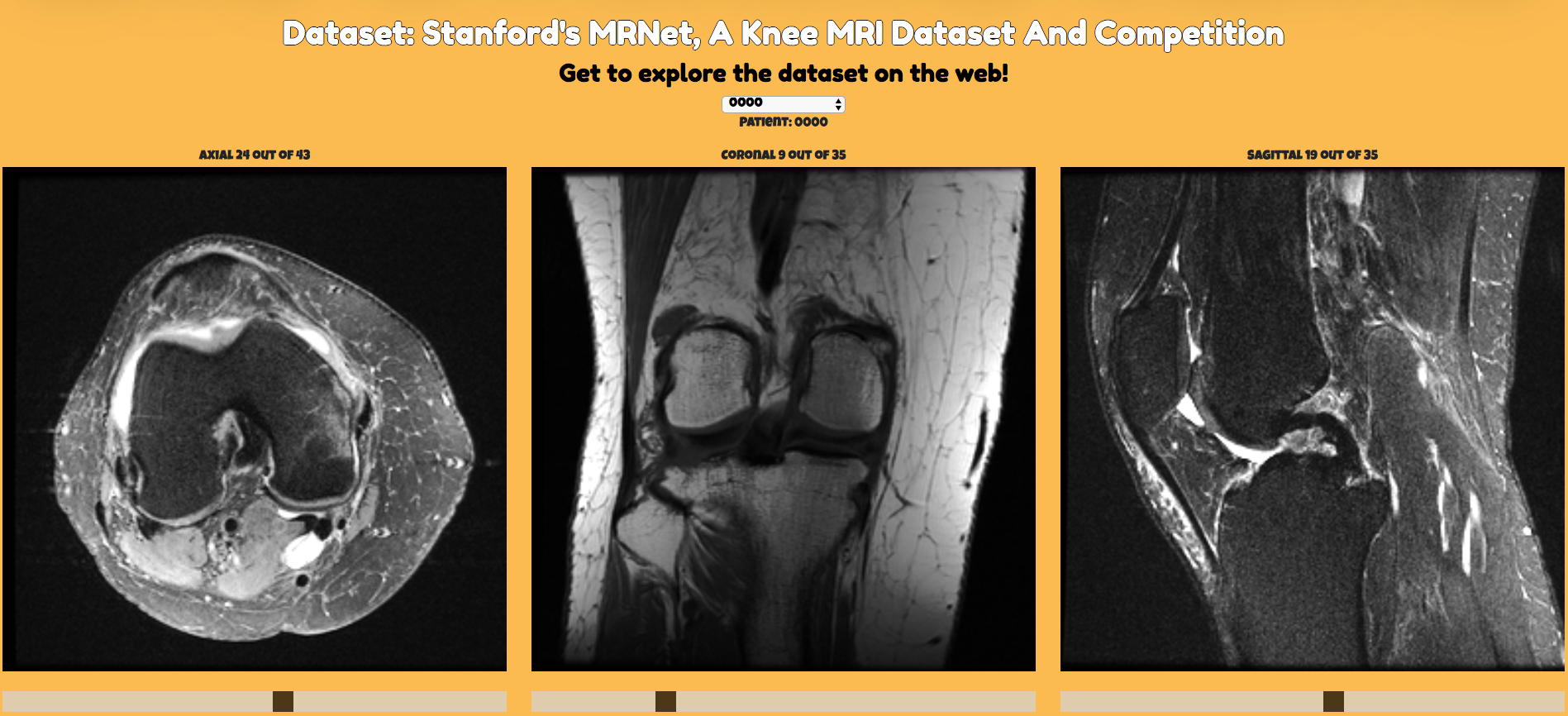}
\caption{Web application developed to explore the MRNet dataset by patient and plane. The application is publicly available in the GitHub repository along with the source code for the models and experiments. See the Acknowledgements~\ref{sec:acks} section. }
\label{fig:web}
\end{figure*}

In order to explore the dataset and after saving the images for each of the planes for all patients, we developed a web application that allowed us to inspect and explore MRI images per plane for any patient. Figure~\ref{fig:web} shows a screenshot from this application. This is similar to Ahmed Besbes' inline Matplotlib and Python tool\footnote{\url{https://www.ahmedbesbes.com/blog/acl-tear-detection-part-1}} but running on the web and using Javascript. Being able to inspect MRIs directly ourselves helped to guide us in choosing some of the machine learning features and configurations we implemented and tested. 

\section{MRNet Architecture and Related Work} \label{sec:mrnet}

Machine learning is one of the buildings blocks of modern artificial intelligence and it has seeped into many applications in almost every domain including heath, transport, finance, entertainment, politics and more. One of the most significant developments in machine learning is the emergence of deep learning \cite{goodfellow2016deep}, a collection of techniques loosely inspired by the architecture of the human brain, which has yielded huge improvements in the effectiveness of tasks like classification and prediction of outcomes.

There has been a lot of work published on using machine learning for analysis of MRIs because of the complexity of the task for a human to perform, the availability of labelled data and the performance of deep learning in this and similar computer vision tasks. A comprehensive survey of the use of deep learning for analysis of MRI images was published in \cite{liu2018applications}, which covered applications of MRI image detection, registration, segmentation and classification. That survey article concentrated mostly on processing MRIs of the (human) brain as well as describing a range of deep learning tools which are available. 

One of the clear messages to emerge from the survey paper \cite{liu2018applications}  is that there are a large number of deep learning architectures that researchers can choose from in implementing a deep learning approach to MRI processing.

Implementing and configuring any effective machine learning application strongly depends on  choosing the right combination of architecture and hyperparameters and this has been the case since machine learning started to become popular nearly 2 decades ago \cite{chapelle2002choosing}.  Optimising hyperparameter choices in deep learning is even more challenging because
the choice of fundamental architecture of the deep network is combined with the many other hyperparameter choices to to be made \cite{denil2013predicting}.
This can be seen in the  exploration of deep learning architectures for predicting DNA- and RNA-binding described in \cite{10.1093/bioinformatics/btz339}.

To push forward the application of deep learning on  MRI data, in 2018 Bien {\em et al.} published the MRI dataset and proposed the MRNet architecture \cite{bien2018deep}, a Convolutional Neural Network (CNN) that predicts, given an MRI, the probability of having either an ACL tear, a meniscus tear, or an abnormal exam on the given knee. An anterior cruciate ligament (ACL) tear is usually a complete tear of the ligament where it has been split into two pieces resulting in an unstable knee. A meniscal tear is a tear of the cartilage that provides a cushion between the bones in the leg, and an abnormal exam is a catch-all classification for these and other knee injuries.

In the work in \cite{bien2018deep} a CNN was trained for each plane: axial, coronal and sagittal, and its three outputs were combined to produce a single probability for each type of knee injury. This process was performed for each particular task: detecting an ACL tear, a meniscal tear or an abnormality. 

The input to MRNet has dimensions $s \times 3 \times 256 \times 256$, where $s$ is the number of images or slices in each plane of the MRI, 3 is the number of the RGB channels and $256 \times 256$ is the dimensionality of the 2D images. We examined the distribution of the number of slices, $s$, a number that varies for all MRIs and this value that can range from 17 to 61 individual 2D images.

MRNet leveraged AlexNet \cite{krizhevsky2012imagenet} as a feature extractor for each 2D image. Each MRI slice or image was passed through AlexNet, obtaining a $s \times 256 \times 7 \times 7$ tensor, then a global average pooling layer was used to reduce the feature vector to $s \times 256$, then to a fully connected layer with a sigmoid activation function to get a probability between 0 and 1. The loss function chosen was binary cross-entropy. The classes are quite imbalanced, as described in Section~\ref{sec:dataset}, so the loss for each example was scaled
inversely proportionally to the number of samples of that class in the dataset. The network was trained using backpropagation and geometric transformations (data augmentation) to improve generalisation and avoid overfitting. Images were rotated randomly, flipped horizontally, and pixels were shifted randomly during training. 

The process described above was carried out for each plane (axial, coronal and sagittal) and models with lowest average loss were retained. A logistic regression was then trained to learn the weights for the different planes and produce a single probability for a particular task: either ACL tear, meniscal tear, or abnormality. 

It is important to highlight a few design decisions on MRNet that are well-explained in \cite{blog}. The number of slices will vary per patient (and also per plane) because of the ways MRI scans are taken by different radiographers and also because of the size of different people's knees, so it is not possible to stack the 2D images in batches and train a network with batch size greater than one. These slices were then input in parallel to AlexNet, which was pre-trained on the ImageNet ILSVRC dataset~\cite{deng2009imagenet}.


\section{Our Proposed Architectures} \label{sec:architecture} 

We propose and we evaluate the performance of the following architectures to train networks and output the probabilities for a patient to have an ACL tear, meniscal tear, or some abnormality on their knee.

\subsection{Training a Deep Residual Network with Transfer Learning}

We leveraged the baseline MRNet architecture and replaced the AlexNet feature extractor and the other layers added on top with a more modern residual architecture \cite{he2016deep} such as Resnet18, Resnet50 and Resnet152. We modified the last layer to output a probability instead of a one-hot softmax vector for a number of classes. We used transfer learning with pre-trained weights from ImageNet. Even when the pre-trained domain is very different, transfer learning still acts as an excellent regulariser and the low-level features learned on the original task appear to work well in practice.

Similar to how MRNet was trained, we repeat the image three times, once per RGB channel, to ensure the input dimensions are compatible with the pre-trained weights from ImageNet. In addition, we input the slices in batches of size one and computed the maximum value of all the slices as the final probability before the back propagation tunes the weights, in the same way as in MRNet. We use cross-entropy loss, the Adam optimizer \cite{kingma2014adam}, and loss is also scaled inversely proportionally to the number of samples of that class in the dataset.

These networks tend to easily overfit so we also applied data augmentation strategies and added a number of new image transformations such as adjusting the contrast of an image by a random factor, applying random gamma adjustment, adjusting the brightness of the image randomly or cropping the image with the center or randomly. A detailed list of the transformations can be found in Table~\ref{tab:policy}. We applied a series of transformations: horizontal flip with probability $p$; one transformation (with the same probability $p$) of the following: random contrast, gamma or brightness; and so on. The clip limit threshold for Contrast Limited Adaptive Histogram Equalization (CLAHE) was two, the other hyperparameters are either specified on the table or the default. The probability $p$ of applying these transformations was optimized on the validation set using grid search with $p$ ranging from 0 to 1 in increments bins of 0.05. 

\begin{table}[!t]
    \centering
    \caption{Data Augmentation Policy. All values correspond to the configuration from 4.1.}
    \vspace{0.1in}
    \label{tab:policy}
    \begin{tabular}{ll}
        \toprule
        \multirow{2}{*}{Sequence} & \multirow{2}{*}{Transformation} \\
        & \\
        \midrule
        Apply & Horizontal Flip \\
        \midrule
        \multirow{3}{*}{Apply one} & Random Contrast \\
        & Random Gamma \\
        & Random Brightness \\
        \midrule
        \multirow{4}{*}{Apply one} & Contrast Limited Adaptive Histogram Equalization \\
        & Sharpen \\
        & Emboss \& Overlay \\
        & Random Brightness Contrast \\
        \midrule
        \multirow{2}{*}{Apply one} & Center Crop \textit{height \& width are 150} \\
        & Random Crop \textit{height \& width are 150} \\
        \bottomrule
    \end{tabular}
\end{table}

Finally, we trained separate models for all three tasks (ACL tear, meniscal tear and abnormality), for each plane (axial, coronal and sagittal), and for each percentage of images being transformed (21 possibilities). Altogether, this was very time and compute intensive but allowed us to explore and discover which are the best-performing parameter combinations. 

\subsection{Training a Deep Residual Network from Scratch \& Use a Fixed Number of Slices}

Instead of repeating each of the 2D images (or slices) three times, one for each RGB color channel, in order to reuse the pre-trained weights, we decided to train a network from scratch using random initialisation of the weights. 

\begin{figure}[htp]

\includegraphics[width=.5\textwidth]{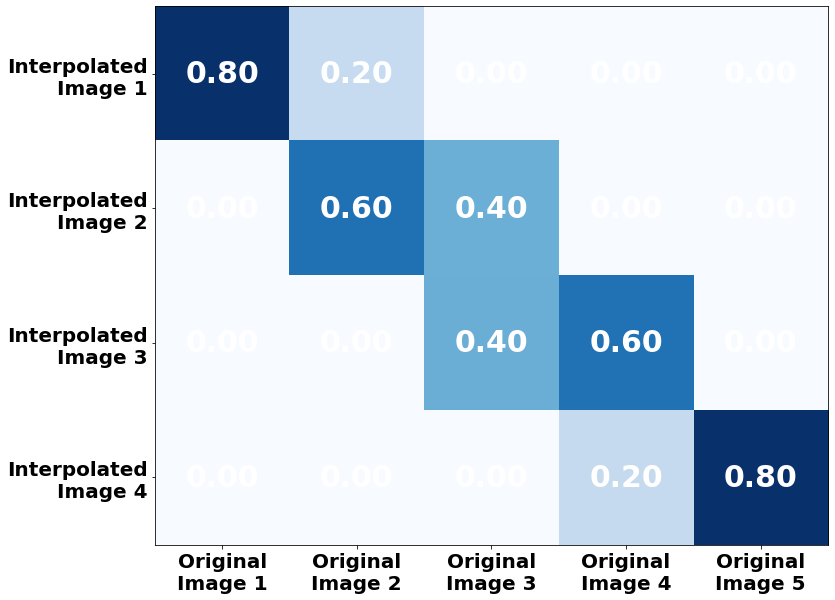}\hfill
\vspace {0.5cm}

\begin{center}
\includegraphics[width=.25\textwidth]{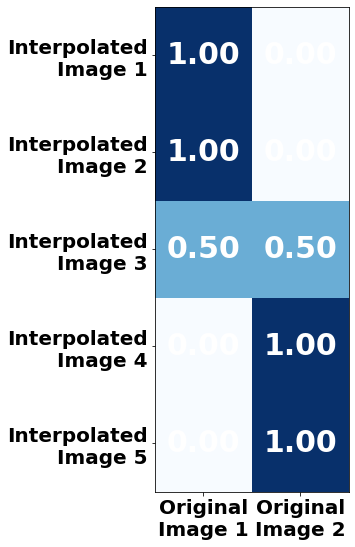}\hfill
\end{center}
\vspace {0.5cm}

\includegraphics[width=.5\textwidth]{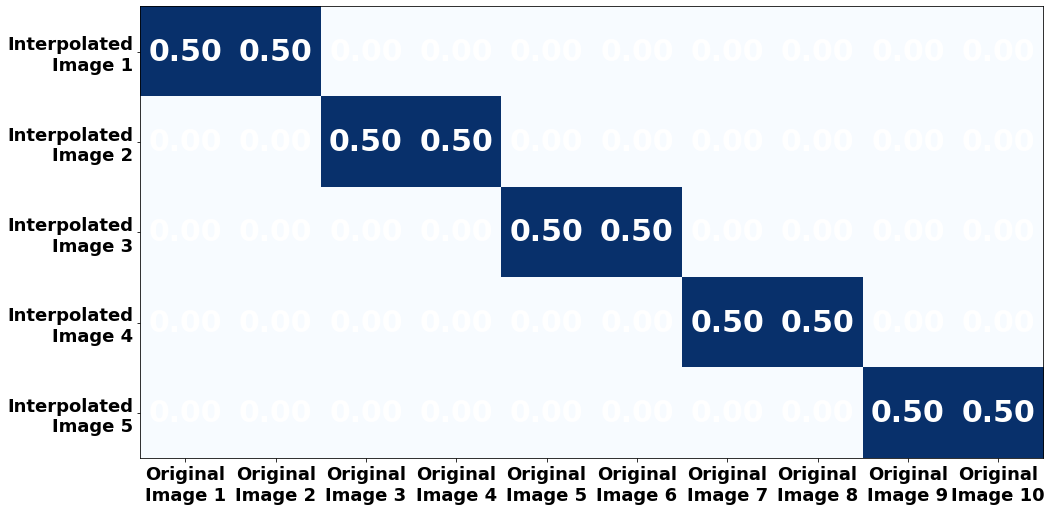}
\caption{Interpolation Examples: Transforming N images into M images. Matrices show the weights to be applied to each original image to interpolate to the newly transformed images}
\label{fig:interpolations}
\end{figure}

In addition, instead of training a network with batch size one and computing the maximum probability of all the slices, we developed methods to input the same number of slices into the network from all patients. We tested the following two approaches:

\begin{enumerate}[leftmargin=*,label=\roman*)]   

\item The first was to use just the middle slices of all planes from all patients at the expense of losing some information. The middle slices are likely to contain more meaningful information than the ones at the beginning and end of the array. The maximum number we could use in this case was 17 slices as that is the minimum number from across all patients. 

\item The second approach was to interpolate up or down to a fixed number of images by applying simple arithmetic operations on the images. Figure~\ref{fig:interpolations} gives a detailed explanation of how to compute the weights to interpolate the images in this way. For instance, in the first example the first image interpolated is a combination of 80\% from the first original image and 20\% from the second original image. We found interpolation performed better than just using the middle slices and the best fixed number to interpolate to was 15 interpolated slices. This was surprising as the performance using larger numbers of images being interpolated was worse. 

\end{enumerate}

Using the approach described in this section, the network  training was performed much faster by not having to repeat the images three times, one per channel, and in addition being able to train them in batches without the need to compute the maximum of all probability predictions. The data augmentation policy was the same as described in the first approach but removing the transformations that needed three channels: Random Brightness, Contrast Limited Adaptive Histogram Equalisation and Random Brightness Contrast. We still computed a probability for each patient and each plane and trained a Logistic Regression classifier to train each task, namely for ACL tear, meniscal tear and abnormality.

As an experiment, we also sliced the set of images or slices horizontally for each patient and each for each task, if we consider that the default slicing technique provided is done vertically. We trained models with those inputs but did not achieve results as good as with the vertical slicing.

\subsection{Training a Multi-Plane Deep Residual Network}

Building from the previous proposed architecture, instead of training three different networks for each of the three different planes (axial, coronal and sagittal) and then combining them with a logistic regression classifier, we also trained one network only whose input was a concatenation of the interpolated slices from the previous approach, that is concatenating the 15 slices per plane which is 45 images as the input. The advantage here is we only had to train one network instead of three and then the logistic regression classifier. The residual network was then trained with 45 channels as the third dimension. Again, this configuration was trained from scratch without transfer learning from ImageNet.

\subsection{Training  a Multi-Plane Multi-Objective Deep Residual Network}

Finally, and also building on the previous proposed architectures, we trained a multi-objective (multi-label) network. In this we trained (from scratch) only one model whose input was  the concatenation of the slices from all planes (45 images) and the output was a vector with the three probabilities for the tasks: ACL tear, meniscal tear and abnormality, at the same time. The motivation for this network was for information learned to predict the probability of an ACL tear or abnormality to aid the prediction of the meniscus task as it was the one with lower AUC.


\section{Results}

Table~\ref{tab:results} gives the experimental results for the methods described in Section~\ref{sec:architecture}. The results from the configuration described in 4.1 show a distinct improvement in average validation AUC (Area Under the Curve) over the baseline in \cite{bien2018deep} and over the test set in Stanford's competition leaderboard\footnote{\url{https://stanfordmlgroup.github.io/competitions/mrnet/}} which is 0.917 (as of October 2020).

\begin{table}[t]
    \centering
    \caption{Results for the proposed architectures. Combined is the accuracy of a logistic regression based ensemble.}
    \vspace{0.1in}
    \label{tab:results}
    \begin{tabular}{lllrr}
        \toprule
        \multirow{2}{*}{Task} & \multirow{2}{*}{Backbone} & \multirow{2}{*}{Plane} & \multirow{2}{*}{Validation AUC} & \multirow{2}{*}{Combined} \\
        & & & & \\
        \midrule
        \multicolumn{5}{l}{\textbf{\textcolor{blue}{Configuration from 4.1}}} \\
        \midrule
        \multirow{3}{*}{ACL} & \multirow{3}{*}{ResNet18} & Axial & 0.9341 & \multirow{3}{*}{0.9557} \\
        & & Coronal & 0.9259 & \\
        & & Sagittal & 0.8967 & 
                \vspace{0.15cm} \\
        \multirow{3}{*}{Meniscus} & \multirow{3}{*}{ResNet18} & Axial & 0.8815 & \multirow{3}{*}{0.9081} \\
        & & Coronal & 0.8521 & \\
        & & Sagittal & 0.7805 &                 \vspace{0.15cm} \
        \\
        \multirow{3}{*}{Abnormal} & \multirow{3}{*}{ResNet18} & Axial & 0.9486 & \multirow{3}{*}{0.9381} \\
        & & Coronal & 0.9128 & \\
        & & Sagittal & 0.9503 & \\
        \midrule
        \multicolumn{4}{l}{Average} & \textbf{0.9340} \\
        \midrule
        \multicolumn{5}{l}{\textbf{\textcolor{blue}{Configuration from 4.2}}} \\
        \midrule
        \multirow{3}{*}{ACL} & \multirow{3}{*}{ResNet18} & Axial & 0.9007 & \multirow{3}{*}{0.9386} \\
        & & Coronal & 0.8681 & \\
        & & Sagittal & 0.7851 &                 \vspace{0.15cm} \
        \\
        \multirow{3}{*}{Meniscus} & \multirow{3}{*}{ResNet18} & Axial & 0.7873 & \multirow{3}{*}{0.8213} \\
        & & Coronal & 0.8012 & \\
        & & Sagittal & 0.7517 &                 \vspace{0.15cm} \
        \\
        \multirow{3}{*}{Abnormal} & \multirow{3}{*}{ResNet18} & Axial & 0.9061 & \multirow{3}{*}{0.9326} \\
        & & Coronal & 0.7288 & \\
        & & Sagittal & 0.9112 & \\
        \midrule
        \multicolumn{4}{l}{Average} & 0.8975 \\
        \midrule
        \multicolumn{5}{l}{\textbf{\textcolor{blue}{Configuration from 4.3}}} \\
        \midrule
        ACL & ResNet18 & 3 Planes  & 0.8883 & N/A \\
        Meniscus & ResNet18 & 3 Planes  & 0.7868 & N/A \\
        Abnormal & ResNet18 & 3 Planes  & 0.8981 & N/A \\
        \midrule
        \multicolumn{3}{l}{Average} & 0.8577 & \\
        \midrule
        \multicolumn{5}{l}{\textbf{\textcolor{blue}{Configuration from 4.4}}} \\
        \midrule
        ACL & \multirow{3}{*}{ResNet18} & \multirow{3}{*}{3 Planes} & 0.8434 & N/A \\
        Meniscus & & & 0.7876 & N/A \\
        Abnormal & & & 0.8522 & N/A \\
        \midrule
        \multicolumn{3}{l}{Average} & 0.8278 & \\
        \bottomrule
    \end{tabular}
\end{table}

\section{Discussion}

From among our proposed deep learning architectures for classification of knee MRIs, the one that performed best was the first: the modified MRNet using residual networks where we trained one model per plane (pre-trained on ImageNet) and then used a logistic regression model to combine the predictions for each task. The trade-off is that this approach takes the longest to train and is the most compute intensive as we feed the same image into the network three times, one per channel, and we calculate a probability per slice. 

Interestingly, the factor that we believe most contributed outperforming the published baseline was the grid search on the percentage of images being augmented. Table~\ref{tab:augmentation} presents details on the percentage of images being augmented for each task and for each plane. For some, such as the meniscus tear task, the percentage of images being augmented is not as high as in the ACL task and having that granularity might have helped get better performance on that task and, hence, get better results overall.

\begin{table}[!t]
    \centering
    \caption{Percentages of images augmented for each task and plane. All values correspond to the configuration from 4.1.}
    \vspace{0.1in}
    \label{tab:augmentation}
    \begin{tabular}{lllr}
        \toprule
        \multirow{2}{*}{Task} & \multirow{2}{*}{Backbone} & \multirow{2}{*}{Plane} & \multirow{2}{*}{Data Augmentation Probability} \\
        & & & \\
        \midrule
        \multirow{1}{*}{ACL} & \multirow{1}{*}{ResNet18} & Axial & 75\% \\
        \cmidrule{3-4}
        & & Coronal & 90\% \\
        \cmidrule{3-4}
        & & Sagittal & 85\% \\
        \midrule
        \multirow{1}{*}{Meniscus} & \multirow{1}{*}{ResNet18} & Axial & 40\% \\
        \cmidrule{3-4}
        & & Coronal & 40\% \\
        \cmidrule{3-4}
        & & Sagittal & 90\% \\
        \midrule
        \multirow{1}{*}{Abnormal} & \multirow{1}{*}{ResNet18} & Axial & 55\% \\
        \cmidrule{3-4}
        & & Coronal & 5\% \\
        \cmidrule{3-4}
        & & Sagittal & 35\% \\
        \bottomrule
    \end{tabular}
\end{table}

For the other proposed approaches, we did not perform grid search but tried some combinations of data augmentation strategies. If we performed this, we anticipate our results would improve slightly, but the relative performance of the approaches would likely stay the same.

The results indicate that the approach using transfer learning (4.1) had a distinct advantage over the other approaches. Direct transfer of weights, however, places architectural limitations on the number of network inputs. The approaches that use a separate network for each plane and task (4.1 and 4.2) give better performance than those trained on a combination of planes (4.3) or using a multi-objective loss (4.4). This is somewhat surprising and indicates some negative transfer among tasks.

\section{Conclusions and Future Work}

In this paper we addressed how to configure a deep learning architecture to improve classification performance on a dataset of MRI images of knees. 
We developed, designed and trained a series of deep learning models to predict the probability of an ACL tear, a meniscus tear or an abnormal exam on a knee given an MRI and we leveraged deep learning and CNN techniques on the MRNet dataset and benchmark. For that, we had images from three planes: axial, coronal and sagittal, as radiologists will typically examine an MRI from different angles. 

Our best-performing deep learning configuration achieved state-of-the-art results by:

\begin{enumerate}[leftmargin=*,label=\roman*)]   
\item leveraging residual networks \cite{he2016deep} and
\item performing fine-grained data augmentation on several planes and tasks.
\end{enumerate}

\noindent 
In addition, we proposed, implemented and evaluated a range of other deep learning architectures that utilise techniques such as interpolation to feed the same number of MRI slices or images to a given network and those:

\begin{enumerate}[leftmargin=*,label=\roman*)]   
\item performed reasonably well 
\item were faster and more efficient to train
\item might help deploy this sort of model in production within constrained environments.
\end{enumerate}

\noindent 

Despite the fact that these other models do not yield AUC performance figures as high as the first model, these approaches should not be overlooked. The ability and flexibility to train and deploy models fast in industry is sometimes more important than small differences in performance.

Ensembling techniques would likely give a boost to the overall performance. We could, for instance, use the models trained with horizontal slicing and ensemble them with the vertical ones. We could also train the same models we proposed with different seeds which would help validate and give the confidence intervals of the performance.

Neural attention mechanisms \cite{vaswani2017attention}, such as the spatial attention mechanism for knee osteoarthritis diagnosis used by Gorriz \emph{et al.}~\cite{pmlr-v102-gorriz19a}, might also help identify which locations in the image the network should focus on and are a promising direction for future work. These have the additional advantage of providing a visual saliency map that can help explain predictions to radiologists.

\medskip

\section*{Acknowledgements} \label{sec:acks}

This research was supported by Science Foundation Ireland under grant SFI/12\-/RC/2289\_P2 and SFI/15/SIRG/3283. Code has been made publicly available on Github at \url{https://github.com/dazcona/mrnet}.

\medskip



\newpage

%

\bibliographystyle{IEEEtran}
\bibliography{mybib}


\end{document}